\documentclass[12pt]{article}

\pdfoutput=1
\usepackage[utf8]{inputenc}
\usepackage[T1]{fontenc}
\usepackage[normalem]{ulem}
\usepackage{verbatim}
\usepackage{amsmath,amssymb,amsthm,amsfonts,mathrsfs,dsfont,braket,array,float,afterpage}
\usepackage{graphicx}
\usepackage[dvipsnames]{xcolor}
\usepackage{slashed}
\usepackage{enumitem}
\usepackage{subfiles}
\usepackage{cite}

\usepackage[colorlinks=false,linkcolor=darkblue,  urlcolor=blue,    filecolor=blue,     citecolor=red,
linktocpage=true,pdfstartview=FitV,bookmarksopen=true,
pdfencoding=auto,
pagebackref]{hyperref}

\usepackage{cleveref}
\usepackage{epsfig}
\usepackage{mathabx}
\usepackage[mathscr]{eucal}
\usepackage{stmaryrd}
\usepackage{comment}
\usepackage{esint}
\usepackage{physics}
\usepackage{setspace}
\usepackage{braket}
\usepackage{float}
\usepackage{url}
\usepackage{mathtools}
\usepackage{bbold}
\usepackage{slashed}
\usepackage{tikz}
\usetikzlibrary{decorations.markings}
\usetikzlibrary{decorations.pathmorphing}
\usepackage{adjustbox}
\usepackage{graphicx}
\usepackage{epstopdf}
\usepackage[labelfont=bf]{caption}
\usepackage{subcaption}
\usepackage{pgfplots}
\usepackage{fancybox}
\usepackage{eurosym}
\usepackage{tcolorbox}
\usepackage{tensor}
\usepackage{soul}
\usepackage{comment}
\usetikzlibrary{arrows.meta}  
\allowdisplaybreaks

\usepackage[margin = 2.37cm]{geometry}
\usepackage[ragged]{footmisc}

\definecolor{orangewolfram}{RGB}{224.584, 155.815, 36.223}
\definecolor{bluewolfram}{RGB}{93.9463, 129.229, 180.998}
\definecolor{greenwolfram}{RGB}{142.846, 176.35, 49.6957}

\def\be{\begin{equation}}
	\def\ee{\end{equation}}

\newcommand{\USp}{{\mathrm{USp}}}

\newcommand{\Spin}{{\mathrm{Spin}}}

\numberwithin{equation}{section}

\newcommand\smallO{
  \mathchoice
    {{\scriptstyle\mathcal{O}}}
    {{\scriptstyle\mathcal{O}}}
    {{\scriptscriptstyle\mathcal{O}}}
    {\scalebox{.7}{$\scriptscriptstyle\mathcal{O}$}}
  }

\pgfplotsset{compat=1.18}

\begin{document}

\begin{titlepage}
\begin{flushright}
SISSA 08/2026/FISI
\end{flushright}
\bigskip
\def\thefootnote{\fnsymbol{footnote}}

\begin{center}
{\LARGE
{\bf Confinement in a finite duality cascade}}
\end{center}

\vskip 5pt

\begin{center}
{\large
Fabrizio Aramini$^{a,b}$, 
Riccardo Argurio$^{c}$, 
Matteo Bertolini$^{a,b}$,
\vskip 4pt
Pietro Moroni$^{a,b}$, Valdo Tatitscheff$^{\,d}$}

\renewcommand{\thefootnote}{\arabic{footnote}}

\vspace{0.25cm}
{\large 
$^a$ {\it SISSA, 
Via Bonomea 265; I 34136 Trieste, Italy\\}
$^b$ {\it INFN - Sezione di Trieste,  
Via Valerio 2, 34127 Trieste, Italy\\}
$^c${\it Physique Th\'eorique et Math\'ematique and International Solvay Institutes \\ 
Universit\'e Libre de Bruxelles, C.P. 231, 1050 Brussels, Belgium\\}
$^d${\it Technical University of Munich, TUM School of Computation, Information and Technology, Department of Mathematics, \\
Boltzmannstraße 3,
85748 Garching bei München, Germany}
\vskip 5pt
{\texttt{faramini,bertmat,pmoroni @sissa.it, Riccardo.Argurio@ulb.be, valdo.tatitscheff@tum.de}}
}
\end{center}

\begin{center} {\bf Abstract} \end{center}
 
\noindent
We provide several consistency checks of confining dynamics in a recently conjectured holographic dual of a four-dimensional ${\cal N}=1$ supersymmetric gauge theory that flows from a conformal manifold in the UV to a finite set of isolated, fully gapped vacua in the IR. This is obtained by considering D3-branes at the conifold singularity in the presence of an O7-plane, leading to a background where all supergravity fields have a non-trivial profile.  We compute holographically the expectation value of a Wilson loop in the fundamental representation and show that it obeys an area law. We then construct the domain walls which interpolate between different vacua in terms of D5-branes wrapping a compact three-cycle of the internal manifold.  Their dynamics is governed by the (2+1)-dimensional ${\cal N}=1$  Yang-Mills-Chern-Simons theory predicted by field theory arguments, that reduces in the deep infrared to a TQFT whose inflow action correctly reproduces the mixed anomaly of the four-dimensional theory. Finally, we argue that, unlike in previous models in the literature, axionic strings are unstable in this background. This implies that the corresponding massless axion that would couple to them is absent, in agreement with the fact that the vacua are fully gapped.

\end{titlepage}
\addtocounter{page}{1}

\setcounter{footnote}{0}

\newpage
\tableofcontents

\section{Introduction}

Confinement is a very interesting physical property of four-dimensional non-abelian gauge theories \cite{Wilson:1974sk, tHooft:1977nqb}. It manifests itself in the most pristine fashion in pure Yang-Mills theories. In such theories, there is a single gapped vacuum (for $\theta\neq\pi$ \cite{Gaiotto:2017yup}). 
The non-trivial low-energy dynamics can be probed by heavy external matter fields that will experience a linearly rising potential, with the slope being a function of their representation under the gauge group. This is equivalent to saying that Wilson loops typically have an area law. The exception is given by  Wilson loops or probes in the adjoint representation, which can be screened by the gauge bosons and hence have a constant expectation value or potential.

The minimally supersymmetric extension of Yang-Mills, namely ${\cal N}=1$ SYM, enjoys much of the same physics, since one adds to the theory only a fermion in the adjoint representation. In addition, the fermion gives rise to a chiral symmetry, which is discrete because of an ABJ-like anomaly and is further spontaneously broken in the confining vacua. As a consequence, there is a finite number of such degenerate vacua \cite{Witten:1982df}, and when spacetime is divided into regions where the theory is in different vacua, there are domain walls between such regions. Such domain walls preserve half of the supersymmetries \cite{Dvali:1996xe}, and must also carry non-trivial topological degrees of freedom to match a mixed 't Hooft anomaly between the discrete axial symmetry and the 1-form symmetry of the gauge sector \cite{Gaiotto:2014kfa}.

All the above expectations are beyond reach with ordinary perturbative techniques in quantum field theory, since they relate to the very low-energy properties of the gauge theories, where their coupling has grown to non-perturbative values. Common approaches to address strongly coupled dynamics are to formulate the theories on a spacetime lattice \cite{Wilson:1974sk}, or to use exact arguments when supersymmetry is present \cite{Seiberg:1994bp}. However, the lattice rarely allows to deduce analytical results, while supersymmetry gives access only to a subsector of the theory (namely, quantities protected by supersymmetry).

Holography \cite{Maldacena:1997re, Witten:1998qj, Gubser:1998bc, Aharony:1999ti}, or gauge-string duality, is a framework where in principle there is analytical access also to non-protected quantities. The difficulty in this approach is to find a string theory set-up which is holographically dual to the gauge theory one wants to study. Furthermore, one would ideally want the string theory set-up to be as much under control as the most celebrated examples, namely those that involve exactly conformal gauge theories (CFTs) on one side and strings in Anti-de Sitter (AdS) on the other.

Holography for confining gauge theories in four dimensions has been addressed soon after its inception \cite{Witten:1998zw, Klebanov:2000hb, Maldacena:2000yy}. All such models display the expected IR behavior, however they all share the problem of having a UV completion (on the QFT side) consisting of an infinite number of degrees of freedom: either in the form of a higher dimensional theory \cite{Witten:1998zw, Maldacena:2000yy}, or of an infinite rank gauge group \cite{Klebanov:2000hb}. On the string theory side, the background is not asymptotically AdS, which makes the extraction of non-protected quantities much less controlled.

A step towards finding a string theory dual of a confining gauge theory that has a better UV behavior was taken in \cite{Argurio:2017upa,Aramini:2025twg}. The set-up of \cite{Aramini:2025twg} is simply an orientifold of the Klebanov-Strassler (KS) set-up \cite{Klebanov:2000hb}. More precisely, one adds a supersymmetry-preserving O7-plane to a stack of regular and fractional D3-branes at the tip of the conifold, that has the effect of turning the SU gauge groups of the KS model into USp ones. The consequences are far-reaching. From the QFT perspective, the characteristic duality cascade becomes such that the difference between the ranks of the two gauge groups decreases towards the UV, until the cascade stops, so that the theory has a UV fixed point with a finite number of degrees of freedom. In the IR, confinement occurs essentially in the same way, except that the vacua are truly gapped, given that USp gauge groups do not allow for a baryonic flat direction, which is instead present in the KS theory. On the string theory side, the supergravity background is built upon those of \cite{Klebanov:2000hb, Benini:2007gx} and now features a running dilaton and a mild curvature singularity at the origin.

In \cite{Aramini:2025twg}, several checks were already performed concerning both the validity of the solution and its matching with field-theory expectations: a precise match of the RG flow through a cascade of Seiberg dualities, and of the beta functions of the gauge groups; a consistency check of the breaking of the R-symmetry to a non-anomalous discrete group; and a match of the $a$ central charge in the UV, together with its monotonicity along the holographic RG flow. It was also shown that the curvature singularity at the origin does not affect any of these computations. Moreover, the dilaton can be made parametrically small, so that along the entire radial range relevant for the description of the RG flow, stringy corrections are suppressed.

In the present paper, we would like to provide further non-trivial checks of the gauge/string dual pair presented in \cite{Aramini:2025twg}, focusing in particular on checks that are pertinent to the confining behavior of the gauge theory. 

In Section \ref{sec: 1-form_WL} we probe confinement in the most direct way by holographically evaluating a Wilson loop in the fundamental representation \cite{Maldacena:1998im, Rey:1998ik}. Since the USp gauge theory at the bottom of the cascade has an unbroken $\mathbb{Z}_2$ 1-form symmetry, we expect the Wilson loop to display area law, or equivalently a linearly rising potential for large distance between probes. We show that this is indeed the case, and we further note that the potential has a transition between nearly conformal to linear behavior, in agreement with field theory expectations. We also highlight that the strings that evaluate holographically the Wilson loop dive into the bulk until a minimal radius that is very close to the origin, but does not reach the latter (unlike in KS), hence completely avoiding the mild curvature singularity.

In Section \ref{sec: domain_wall} we probe the breaking of the discrete R-symmetry by studying the holographic realization of domain walls between different degenerate vacua. As in KS, they are realized by D5-branes wrapping the 3-cycle in the geometry (which is the one of an orientifolded deformed conifold) whose volume remains finite at the origin. We are able to determine that the theory on the effective 3d world-volume at low-energies is a supersymmetric Chern-Simons (CS) theory that is exactly what one expects from arguments based both on supersymmetry, and on the matching of the mixed anomaly between the R-symmetry and the 1-form symmetry.

Finally, in Section \ref{sec: mass_gap} we present a simple argument that shows that, in our setup, the moduli space of the KS solution associated to the baryonic branch \cite{Gubser:2004qj} is completely lifted by the orientifold projection. This agrees with the expectation from the dual gauge theory, where the vacuum described by our solution is fully gapped \cite{Aramini:2025twg}. We conclude the paper with a discussion in Section \ref{sec: disc} of some subtler features related to global aspects of the gauge theory, which require more advanced mathematical tools to analyze the geometry of our setup.

The relevant features of both the gauge theory and of the dual supergravity solution of \cite{Aramini:2025twg} are reported in Appendix \ref{app: SUGRA solution}. Appendices \ref{app: asymptotic expansion} and \ref{app: tension of domain walls} contain the derivation of several analytical results quoted in the main body of the paper.

\section{One-form symmetry and holographic Wilson loops}
\label{sec: 1-form_WL}

In the vacuum that our supergravity solution describes, the theory confines. Hence, we expect the $\mathbb{Z}_2$ one-form symmetry to be preserved and the Wilson loops in the fundamental representation to follow an area law.\footnote{ Because of the presence of matter in the bifundamental representation, at every step of the cascade the unbroken one-form symmetry acts diagonally with respect to the center symmetry of the two symplectic gauge groups, hence it is $\mathbb{Z}_2$. At the last step of the cascade, it reduces to the $\mathbb{Z}_2$ one-form symmetry of the single node $\USp(2N)$ SYM theory. Therefore, the one-form symmetry is $\mathbb{Z}_2$ all along the RG-flow.} In the following, we will test this prediction holographically. Throughout the whole paper, we will work in units of $\alpha'=1$. 

We use the standard dictionary of \cite{Maldacena:1998im,Rey:1998ik} (see also \cite{Brandhuber:1998er, Sonnenschein:1999if}). The novelty is in the non trivial dilaton profile: as we will show below, the dilaton modifies the geometry, preventing the string from penetrating it beyond a finite distance $\tau_\star>0$, even in the limit of infinite length of the Wilson loop. 

The expectation value of a fundamental Wilson loop \
\begin{equation}
    \langle W_\gamma\rangle =\left\langle \Tr\mathcal{P}e^{i\int_\gamma A}\right\rangle
\end{equation}
can be computed holographically as the area of a fundamental string world sheet $\Sigma$ that terminates at the boundary on the contour $\gamma$ 
\begin{equation}
    S=\frac{1}{2\pi} \int_\Sigma \sqrt{\det g}\ .
\end{equation}
Here $g$ is the pullback of the spacetime metric in the string frame. Usually, supergravity solutions are written in the Einstein frame. The metric tensors in the two frames are related by a dilaton dependent conformal transformation, which, in 10 dimensions, reads
\begin{equation}
    g=e^\frac{\phi}{2} g^E\ ,
\end{equation}
where $\phi$ is the dilaton. Consider a 10-dimensional metric of the form
\begin{equation}\label{eq: metric}
    d s^2_{10}=h^{-1/2}(r) d x^2_{1,3}+h^{1/2}(r) d s^2_6 \ .
\end{equation}
Both in our case and in the warped deformed conifold background of \cite{Klebanov:2000hb}, the six-dimensional space with metric $d s^2_6$ is not a cone over a compact space $X_5$ (because of the deformation). Nevertheless, we can write it as
\begin{equation}
    d s^2_6= d r^2+ r^2 d X_5^2(r) \ ,
\end{equation}
where $d X_5$ depends in a complicated way on all six internal coordinates, see eq.~\eqref{eq: metric Ansatz}. 

We consider a string with world-sheet coordinates $(\sigma_1,\sigma_2)$, sitting at fixed angular coordinates and localized at fixed $y$ and $z$ in $\mathbb{R}^{1,3}$. We choose a gauge such that $\sigma_1$ is aligned along the time direction $t$ while $\sigma_2$ describes a one-dimensional slice in the two-dimensional space spanned by $r$ and $x$. We pick $x=\sigma_2$, which implies, in turn, that $r=r(\sigma_2)=r(x)$.

The pullback of the metric \eqref{eq: metric} onto the world-sheet is then
\begin{equation}\label{eq: pullback metric}
    h^{-\frac{1}{2}} (r) (d t^2+d x^2) + h^{\frac{1}{2}} (r) \, d r^2 = h^{-\frac{1}{2}} (r)\left(d t^2+(1+h(r)r'^{\, 2}) \, d x^2)\right) \ ,
\end{equation}
where $r' = d r/d x$ and the boundary is at $r \rightarrow +\infty$. This gives the Nambu-Goto (NG) action
\begin{equation}
\label{NG_1}
    S=\frac{T}{2\pi} \int_{-\frac{L}{2}}^{\frac{L}{2}} \frac{e^{\frac{\phi}{2}}}{h^{\frac{1}{2}}}{}\sqrt{1+hr'^2} \, d x \ ,
\end{equation}
where we already integrated in $t$ and chose the endpoints of the string to be at $x=-\frac{L}{2}$ and $x=\frac{L}{2}$, so that the distance between the probe quarks is $L$ and the embedding is symmetric with respect to $x=0$. 

The action is that of a dynamical system where the role of time is played by $x$ and that of the coordinates $q,\, \dot q$ by $r$ and $r'$. Since the Lagrangian $\mathcal{L}$ is independent of $x$, the ``energy'' is conserved on-shell, i.e.
\begin{equation}\label{eq:energy}
    \frac{\partial \mathcal{L}}{\partial r'} \, r'- \mathcal{L} = -\frac{e^{\frac{\phi}{2}}}{h^{\frac{1}{2}}\sqrt{1+hr'^{\, 2}}} = \mathrm{const.} \ ,
\end{equation}
where the constant is determined observing that $r'=0$ at the minimal value of the radius, that we call $r_m$. This tells us that
\begin{equation}\label{eq:r'}
    \sqrt{1+hr'^{\, 2}} = \frac{h_m^{\frac{1}{2}}e^{\frac{\phi}{2}}}{ h^{\frac{1}{2}}e^{\frac{\phi_m}{2}}}\ ,
\end{equation}
where we have set $\phi(r_m) = \phi_m$ and $h(r_m) = h_m$ to simplify the notation.
Substituting this back into eq.~\eqref{NG_1} gives
\begin{equation}\label{eq:NG}
     S =  \frac{T}{\pi } \int_0^{\frac{L}{2}} \frac{h_m^{\frac{1}{2}}e^{\phi}}{ h\, e^{\frac{\phi_m}{2}}} \,d x \ ,
\end{equation}
where now the action is on-shell (the different solutions to the equations of motion are given by different values of $r_m$ which, in turn, correspond to different values of the boundary condition $L$). In deriving this expression, we also used the symmetry of the problem under $x\rightarrow -x$. We can furthermore use \eqref{eq:r'} to get a differential equation for $r$ as a function of $x$ 
\begin{equation}\label{eq:diff eq r'}
    \sqrt{1+hr'^{\, 2}} = \frac{h_m^{\frac{1}{2}}e^{\frac{\phi}{2}}}{ h^{\frac{1}{2}}e^{\frac{\phi_m}{2}}} \implies hr'^{\, 2} = \frac{h_me^{\phi}}{ h\, e^{\phi_m}}-1 \implies r'=\pm \frac{1}{h^{\frac{1}{2}}} \sqrt{\frac{h_me^{\phi}}{ h\ e^{\phi_m}}-1} \ .
\end{equation}
The radial coordinate $r$ is an increasing function of $x\in[0,L/2]$, where $r(0)=r_m$ and $\lim_{x\rightarrow L/2} r(x) = +\infty$. This means that we have to select the equation with positive sign above. 

Integrating \eqref{eq:diff eq r'} with respect to $x \in [0,\frac{L}{2}]$, one easily obtains 
\begin{equation}\label{eq:L/2}
    L = 
    2\frac{e^{\frac{\phi_m}{2}}}{h_m^{\frac{1}{2}}} \int_{r_m}^{\infty} \frac{h\ e^{-\frac{\phi}{2}}}{\sqrt{1-\frac{h\, e^{\phi_m}}{h_m e^\phi}}} \, d r \ ,
\end{equation}
which is an equation for $r_m$ as a function of $L$. We can also use \eqref{eq:diff eq r'} to change the integration variable in \eqref{eq:NG}, obtaining
\begin{equation}\label{eq:final_NG}
    S = \frac{T}{\pi } \int_{r_m}^{\infty} \frac{e^\frac{\phi}{2}}{\sqrt{1-\frac{h\, e^{\phi_m}}{h_m e^\phi}}} \, d r \ ,
\end{equation}
which is the on-shell NG action as a function of $r_m$. From eqs.~\eqref{eq:L/2} and \eqref{eq:final_NG} one can then obtain $S$ as a function of $L$, and determine the potential between probe quarks. 

One subtlety regarding eq.~\eqref{eq:final_NG} is that it is in fact divergent. This property is the supergravity counterpart of the fact that the rest energy of the probe quarks in the dual gauge theory gives an infinite contribution to the expectation value of the Wilson loop. Since what matters is the interaction energy, rather than the total energy of the system, one must subtract the contribution from the rest energy of the quarks, which is given by the NG action for strings that are straight lines at $x=\pm \frac{L}{2}$. This finally yields
\begin{equation}\label{eq:S regularized}
    S = \frac{T}{\pi } \int_{r_m}^{\infty} e^\frac{\phi}{2}\left(\frac{1}{\sqrt{1-\frac{h\, e^{\phi_m}}{h_m e^\phi}}} -1 \right)\, d r -  \frac{T}{\pi } \int_{0}^{r_m} e^\frac{\phi}{2}\, d r \ ,
\end{equation}
which is convergent. 

Up to this point, the discussion has been general and applies to any metric of the form \eqref{eq: metric}. For solutions with constant dilaton, such as the KS solution, the dilaton contributes only an overall inessential factor. In these cases, the string penetrates more and more into the bulk of the geometry as one increases the length of the Wilson loop, and one can prove that $r_m \to 0$ as $L \to \infty$. In the following, we instead specialize to our background with non-constant dilaton in order to illustrate how the situation differs.

The explicit expressions of the warp factor and of the dilaton simplify if we use a different radial coordinate $\tau$, defined by $d \tau= 3 e^{-G_3} dr$ where 
\begin{equation}
    e^{2G_3 (\tau)}= 6 \mu^{\frac{4}{3}} \frac{\tau-\tau_0}{\tilde{\Lambda}^2(\tau)}\ , \quad \tilde{\Lambda}(\tau)=\frac{\bigg(2\left(\sinh{(
    2 \tau)}-\tau\right)(\tau-\tau_0)-\cosh \left(2\tau\right)+2\tau\tau_0+1\bigg)^\frac{1}{3}}{\sinh{\tau}} \ ,
\end{equation}
with $\tau_0<0$ an integration constant. With this change of coordinates, $S$ and $L$ become
\begin{equation}\label{eq:S and L}
\begin{split}
    S &= \frac{T}{3 \pi } \int_{\tau_m}^{\infty} \left( \frac{1}{{\sqrt{1-\frac{h\, e^{\phi_m}}{h_m e^\phi}}}} -1 \right) e^{G_3} e^\frac{\phi}{2}\, d\tau - \frac{T}{3 \pi } \int_0^{\tau_m} e^{G_3} e^\frac{\phi}{2}\, d\tau\ , \\
    L &=  \frac{2 e^\frac{\phi_m}{2}}{3h_m^\frac{1}{2}} \int_{\tau_m}^{\infty} \frac{h}{\sqrt{1-\frac{h\,  e^{\phi_m}}{h_m e^\phi}}} e^{G_3} e^{-\frac{\phi}{2}}\, d \tau \ ,
\end{split}
\end{equation}
where $\tau_m = \tau (r_m)$. The explicit expressions for the warp factor and the dilaton as a function of $\tau$ are given in Appendix \ref{app: SUGRA solution}, eqs. \eqref{eq: solution deformed conifold}-\eqref{eq: dilaton_1}. The warp factor $h(\tau)$ is a decreasing function of $\tau$, thus, with a constant dilaton, the argument of the square roots in \eqref{eq:S and L} is positive $\forall \tau \ge \tau_m, \forall\tau_m \ge 0$. However, in our case the dilaton has a non trivial profile and is a decreasing function as well, so the above does not necessarily hold true anymore. 

More precisely, $e^{-\phi}$ is a linear, monotonically increasing function of $\tau$, see eq.~\eqref{eq: dilaton_1}, while $h(\tau)$ has the following expansion for small and large $\tau$
\begin{equation}\label{eq:expansion h}
h(\tau)\underset{\tau\to 0}{\sim} h_0-h_2 \tau^2 +\mathcal{O}(\tau^3)\ , \quad h(\tau)\underset{\tau\to \infty}{\sim} h_\infty \,\tau^{-\frac{2}{3}}e^{-\frac{4}{3}\tau}(1+\mathcal{O}(1/\tau))\ ,
\end{equation}
with $h_0, h_2$ and $h_\infty$ positive ($\tau_0$-dependent) constants.

The function $he^{-\phi}$ has a unique maximum for every $\tau_0<0$, as can be verified by plotting it for different values of $\tau_0$ and by analyzing its behavior in the limits of small and large $\tau_0$. Note that the validity of the semi-classical approximation requires the string coupling to be small, which, by eq.~\eqref{eq: dilaton_1}, is equivalent to $|\tau_0|\gg1$. In the following, we denote the position of this maximum by $\tau=\tau_\star$ and estimate it in the regime $|\tau_0|\gg1$. To do so, we first assume that $\tau_\star$ lies in the regime where the small-$\tau$ expansion is valid, and subsequently provide an a posteriori check of this assumption. 
It is clear that $\tau_m$ cannot be smaller than $\tau_\star$, otherwise for $\tau_m<\tau<\tau_\star$ we would have $\frac{h\, e^{\phi_m}}{h_m e^\phi}>1$ and a negative argument for the square root in both expressions for $S$ and $L$. Hence, $\tau_\star$ is the minimum value of the radial variable that a string can reach, meaning that $\tau_m\to\tau_\star$ as $L\to\infty$, as shown in Appendix \ref{app: asymptotic expansion}. Let us evaluate $\tau_\star$ by imposing $(he^{-\phi})'=0$ in the range of small $\tau$. We get
\begin{equation}
\label{eq:taustar_1}
    \tau_\star \sim \frac{h_0}{2h_2 |\tau_0|}\ .
\end{equation}
 Both $h_0$ and $h_2$ depend on $\tau_0$, and one can easily show that $h_2 \ \propto \ |\tau_0|^{-\frac{5}{3}}$, while $h_0 \sim |\tau_0|^{-\frac{5}{3}}$ for $|\tau_0| \rightarrow \infty$, so that $h_0 \sim h_2$ for large $|\tau_0|$. From \eqref{eq:taustar_1} we can then conclude that $\tau_\star$ is indeed very close to the tip, where the small $\tau$ expansion of $h$ is valid, as anticipated. 

Let us emphasize that a string that terminates on an infinitely long Wilson loop will probe the geometry only up to $\tau_m=\tau_{\star}>0$. This in particular means that such strings naturally avoid the tip $\tau=0$ which is the locus of a (mild) curvature singularity, see \cite{Aramini:2025twg}.

\begin{figure}
    \centering
    \includegraphics[width=0.7\linewidth]{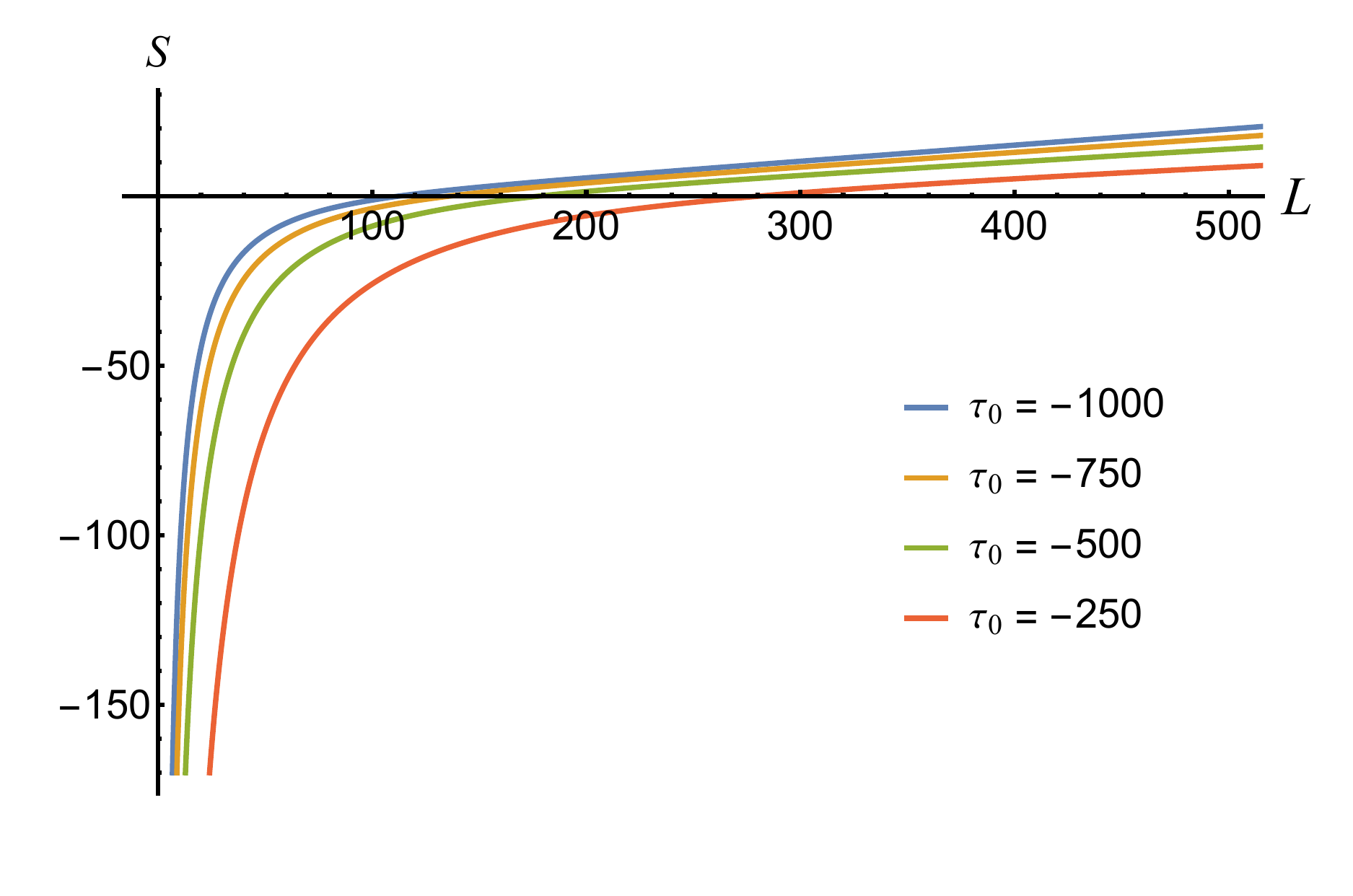}
    \caption{$S$ as a function of $L$ for different values of $\tau_0$. The plot shows linear behavior at large $L$ and conformal-like behavior at small $L$.}
    \label{fig:S(L)}
\end{figure}

The dependence of $S$ on $L$ as $\tau_m$ varies in the range $(\tau_{\star},\infty)$ can be studied numerically and is plotted in figure \ref{fig:S(L)}. Importantly, one sees that $S$ becomes linear in $L$ for large $L$, signaling that the Wilson loop follows an area law   and confirming that our background is dual to a confining vacuum, with unbroken $\mathbb{Z}_2$ one-form symmetry. The linear behavior can also be checked analytically with an asymptotic expansion for $\tau_m \sim \tau_{\star}$. Similarly, one also finds that, for small $L$, $S \sim - 1/(L\sqrt{\log (1/L)})$. This behavior is also expected, as in the far UV the field theory reaches a fixed point \cite{Aramini:2025twg}, but it does so with a running gauge coupling, so that the conformal behavior $S \sim -  1/L$ is distorted. Some details on both asymptotic expansions are presented in Appendix \ref{app: asymptotic expansion}. In figures \ref{fig:L(tau)} and \ref{fig:S(tau)} the length of the string and the NG action are plotted as functions of $\tau_m$. They both diverge  for  $\tau_m=\tau_{\star}$, which is very close to $0$ for large $|\tau_0|$. The estimate \eqref{eq:taustar_1} can be checked in figure \ref{fig:hephi(tau)}, where $(he^{-\phi})'$ is plotted against $\tau_m$.

\begin{figure}[!htbp]
    \centering
    \begin{subfigure}{0.49\textwidth}
        \centering
        \includegraphics[width=\textwidth]{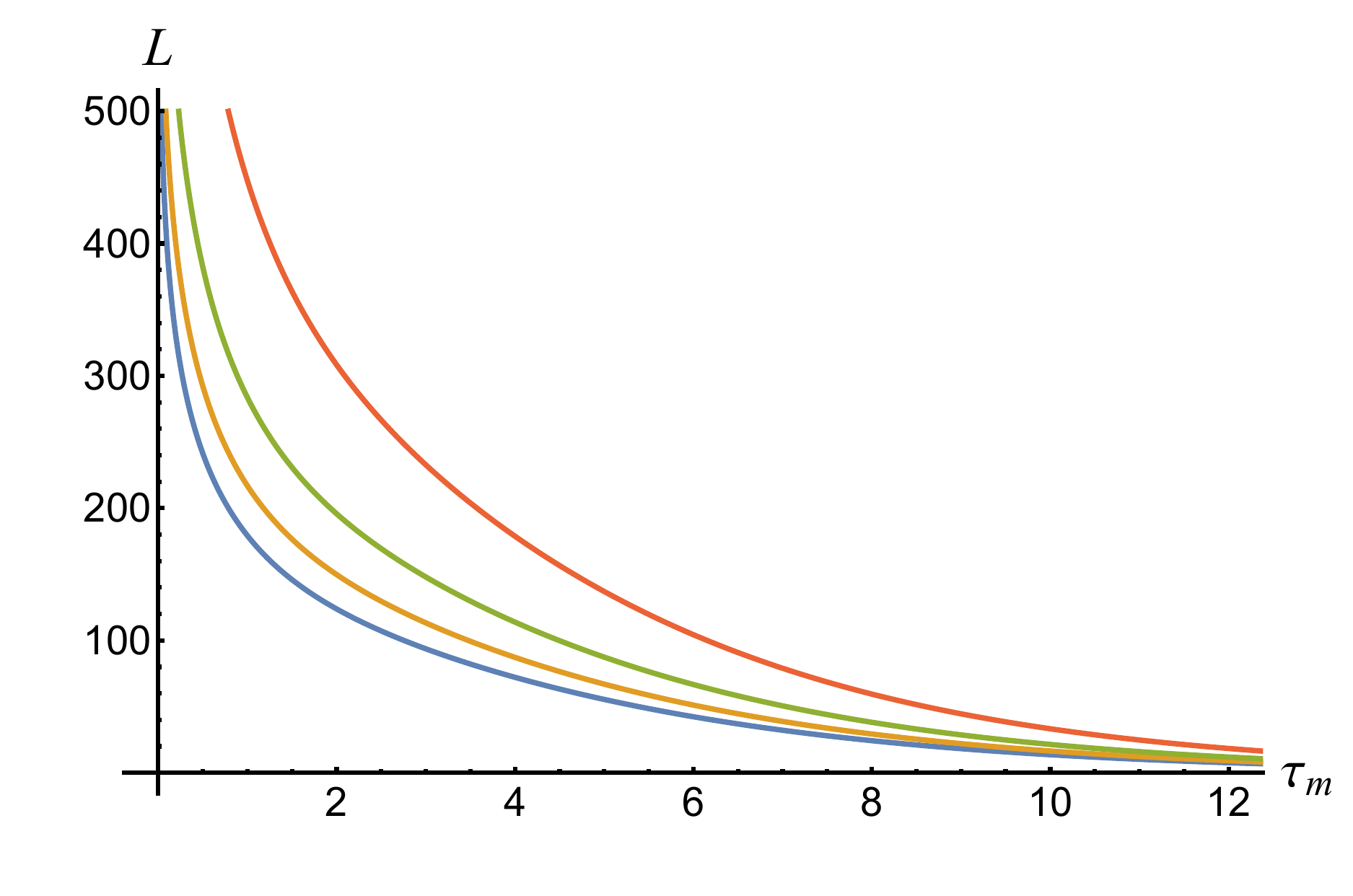}
        \caption{$L$ as a function of $\tau_m$.}
        \label{fig:L(tau)}
    \end{subfigure}
     \hfill
     \begin{subfigure}{0.49\textwidth}
        \centering
        \includegraphics[width=\textwidth]{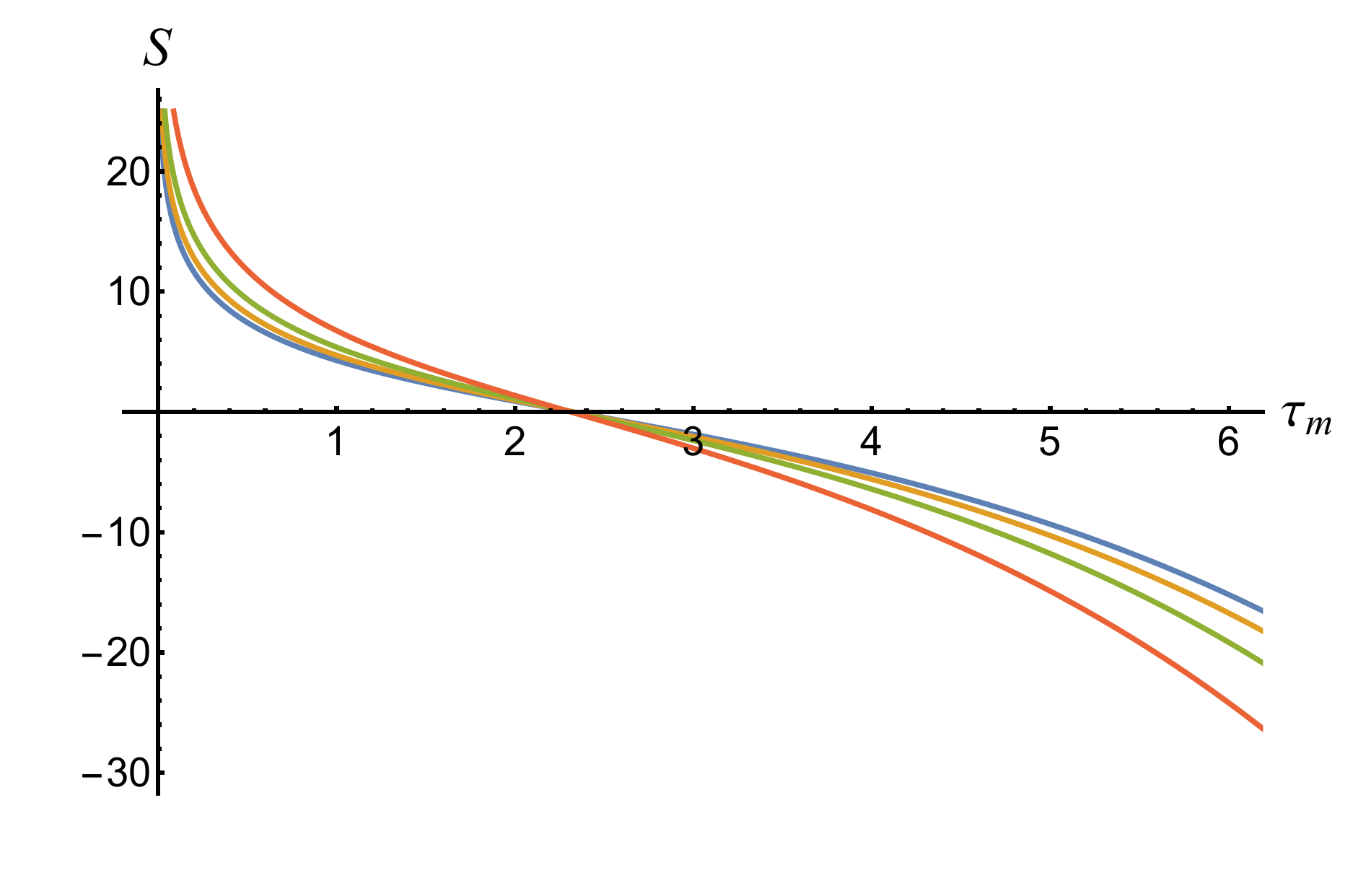}
        \caption{$S$ as a function of $\tau_m$.}
        \label{fig:S(tau)}
    \end{subfigure}
     
    \vspace{0.6cm}

    \begin{subfigure}{0.6\textwidth}
        \centering
        \includegraphics[width=\textwidth]{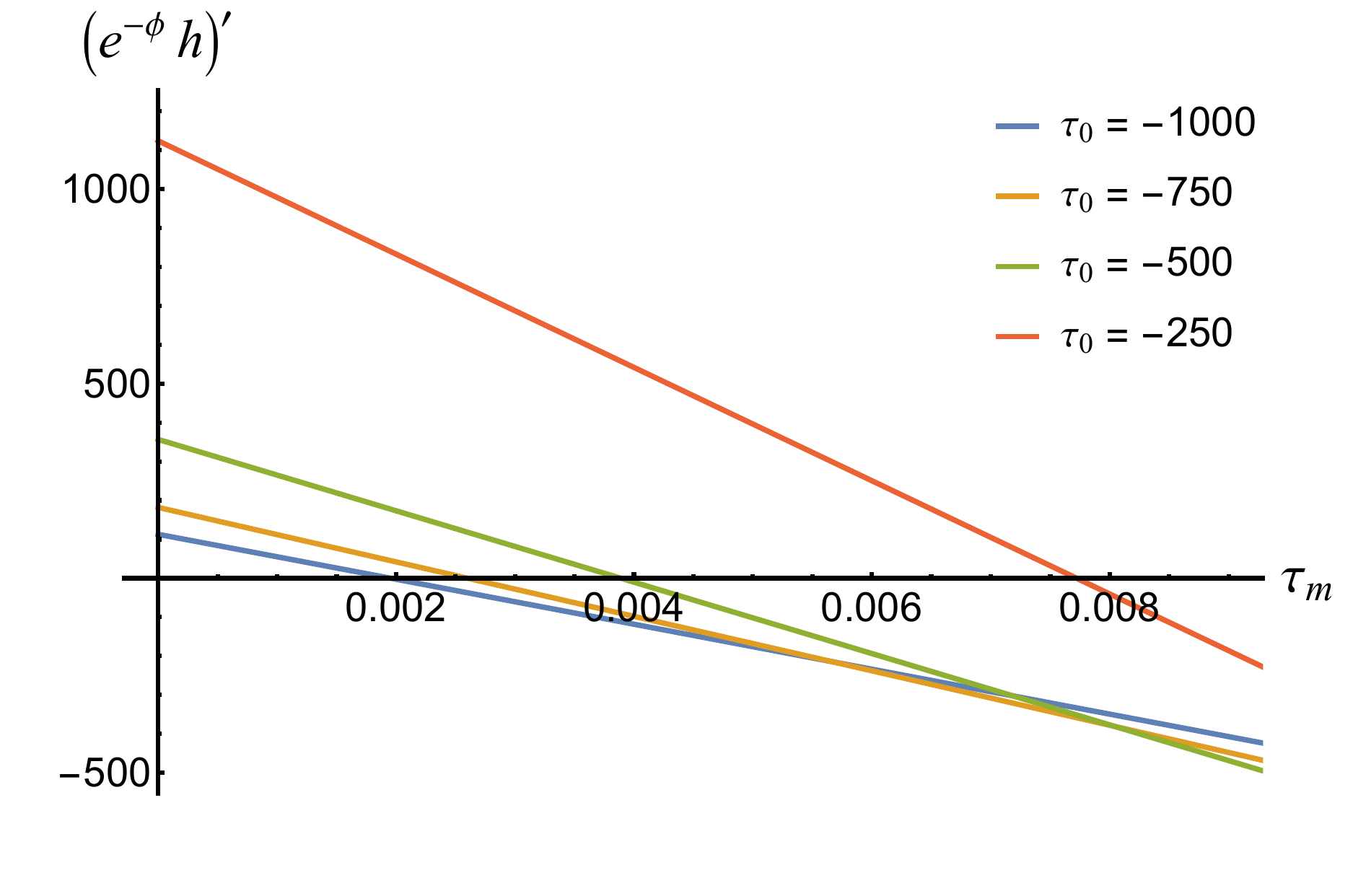}
        \caption{$(he^{-\phi})'$ as a function of $\tau_m$. The zero agrees with the estimate \eqref{eq:taustar_1}.}
        \label{fig:hephi(tau)}
    \end{subfigure}
    \caption{We show the plot of $L$ (resp. $S$) against $\tau_m$ in fig. \ref{fig:L(tau)} (resp. \ref{fig:S(tau)}) for different values of $\tau_0$. Both functions diverge at $\tau_m=\tau_\star$, which is the zero of the function $(e^{-\phi}h)'$ plotted against $\tau_m$ in fig. \ref{fig:hephi(tau)}.}
    \label{fig:three_panels} 
\end{figure}

\section{Domain walls}
\label{sec: domain_wall}

In the vacuum that our supergravity solution describes, the orientifold quiver gauge theory reduces to pure SYM with gauge group $\USp(2N)$, which has a non-ABJ-anomalous $\mathbb{Z}_{2(N+1)}$ R-symmetry and a $\mathbb{Z}_2^{(1)}$ 1-form symmetry. 

This theory confines. In the confining vacua the 1-form symmetry is preserved while the R-symmetry is spontaneously broken to $\mathbb{Z}_2$ by gaugino condensation. Therefore, we expect $N + 1$ gapped vacua. Since they are rotated into each other by the broken R-symmetry, these vacua are expected to be physically equivalent.

For any pair of vacua, one can construct field configurations in which the theory sits in two different vacua on the left and right half-spaces, respectively. In such configurations, a domain wall (i.e. a tensionful codimension-one solitonic object) must necessarily separate the two spatial regions. We would like to construct these (BPS) domain walls holographically.

\subsection{Domain walls - field theory}

From the field theory perspective, a minimal requirement on the 3d theory dressing the domain wall (DW) is due to the following.
$\USp(2N)$ SYM has a mixed anomaly between the 1-form symmetry and the (discrete) 0-form R-symmetry \cite{Cordova:2019uob, Argurio:2023lwl}. The mixed anomaly is captured by the appearance of the following phase when one performs $\theta\to\theta+2\pi$ (i.e.~one unit of discrete R-symmetry transformation) in presence of a background field for the 1-form symmetry
\begin{equation}
\label{anomaly usp}
{\cal A} = \exp\left(2 \pi i \frac{N}{2}\int  \frac{{\cal P}(B_2)}{2} \right) \ ,
\end{equation}
where the integral is over the four-dimensional space ${\cal M}$ where the gauge theory lives, $B_2$ is an integral 2-cocycle background gauge field for $\mathbb{Z}_2^{(1)}$characterized up to gauge transformations by its class in $H^2({\cal M}, \mathbb{Z}_2)$ that we also denote by $B_2$, and ${\cal P}(B_2)$ its Pontryagin square.\footnote{The Pontryagin square is the natural square operation on discrete gauge fields, a map between $H^2({\cal M},\mathbb{Z}_2)$ and $H^4({\cal M},\mathbb{Z}_4)$, with $\frac{1}{2} \int {\cal P}(B_2) \in \mathbb{Z}_2$ in the present case where ${\cal M}$ is taken to be a manifold with a spin structure.} Such a phase is commonly taken to represent a symmetry protected topological (SPT) phase, namely a trivially gapped phase that nevertheless responds to changes in the background gauge field of an unbroken symmetry. 

This anomaly should be captured by the IR physics. In particular, when going from one of the $N+1$ vacua to the next one, $\theta$ shifts precisely by $2\pi$ and hence one potentially generates a non-trivial SPT phase.

Notice that since the integral of the Pontryagin square is even on spin manifolds, the anomaly \eqref{anomaly usp} becomes trivial for $N$ even. This implies that for even $N$ the vacua are all in the same SPT phase. On the contrary, when $N$ is odd, the anomaly is valued in $\mathbb{Z}_2$ and the vacua alternate between the two distinct SPT phases. Nicely, this is consistent with the number of vacua: when $N$ is odd, there is an even number of vacua, so that they can alternate between two different phases. This is clearly not possible if the number of vacua is odd, consistently with the absence of anomaly for even $N$.

A consequence of two vacua being in different SPT phases is that the domain wall interpolating between the two has to carry a non-trivial TQFT on its world-volume in order to compensate the change in the SPT. In other words, its three-dimensional world-volume theory must have an anomaly such that its four-dimensional inflow is precisely the SPT phase. We are thus expecting that, at least in the odd-$N$ case, the DW theory has a $\mathbb{Z}_2$ 1-form symmetry with an anomaly given by $N\! \mod 2$.

However, because of supersymmetry, we can be much more precise. 
In ${\cal N}=1$ $\USp(2N)$ SYM the dynamics of a $k$-wall\footnote{A $k$-wall is a domain wall interpolating between the $j^{th}$ vacuum and the $(j+k)^{th}$ vacuum, with $j=1,..., N+1$.} is described by a (2+1)-dimensional YM-CS theory \cite{Bashmakov:2018ghn} (for an alternative but IR equivalent description see \cite{Delmastro:2020dkz}) 
\begin{equation}
\label{dw_sp_1}
{\cal N}=1 \ \, \USp(2k)_{N+1}\ \, \text{coupled to a rank-2 antisymmetric scalar multiplet} \,\, \Phi \ .
\end{equation}
The antisymmetric representation is reducible and splits as the direct sum of the singlet representation, proportional to the defining symplectic matrix $\Omega$, and the $\Omega$-traceless antisymmetric one. The former describes the Goldstone mode associated to broken translations and describes the center-of-mass motion of the domain wall perpendicular to its worldvolume. The latter parametrizes the classical moduli space. However, quantum corrections lift all flat directions, generating a negative mass around $\Phi=0$. This implies that if one is interested in the domain wall IR physics, this field should be integrated out. 
The same holds for the pure gauge sector since the Chern-Simons part of the (super) YM-CS action induces a (classical) mass term both for the gauge bosons and the gaugino. 
This implies that at energies below their mass they should be integrated out.
Therefore, the domain wall theory is a pure TQFT in the deep IR. 

Both the gaugino and the fermion in the scalar superfield are Majorana, whose mass term is parity odd, and integrating them out shifts the CS level \cite{Redlich:1983dv}. Taking into account that the Dynkin index of the Anti-symmetric and Adjoint representations are, respectively, $T(AS) = k-1$ and $T(Adj) = k+1$ one easily gets, from \eqref{dw_sp_1}, that the domain wall  in the deep IR is described by 
\begin{equation}
\label{dw_sp_3}
\USp(2k)_{N+1 - k}\ 
\end{equation}
pure CS theory. 
Notice that for $k=N+1$ the CS term vanishes and the TQFT becomes trivial, consistently with the fact that in this case one is describing a domain wall interpolating between a vacuum and itself, and would hence be unstable towards decay into the 4d vacuum.

Choosing $k=1$ one describes a 1-wall, that is, a domain wall interpolating between a vacuum and its nearest neighbor. In this case the scalar multiplet has just the singlet component, which is completely decoupled, and the domain wall theory is simply 
\begin{equation}
\label{purek1_Susy}
 {\cal N}=1 \ \, \USp(2)_{N+1}\ .
\end{equation}
In the IR, upon integrating out the gauge sector, this becomes 
\begin{equation}
\label{purek1}
\mathrm{SU}(2)_N \ \text{pure CS}\ ,
\end{equation}
where we used the fact that $\USp(2) \equiv \mathrm{SU}(2)$, in our conventions. 

The theory \eqref{purek1} has a 1-form symmetry generated by all its non-trivial topological lines. In general, the lines form a non-trivial fusion category which is not a group. However, if we focus on the subset of lines that remain topological in presence of a Yang-Mills UV regularization, they generate an abelian 1-form symmetry group $\mathbb{Z}_2$ (namely, there is essentially one non-trivial line besides the identity). This symmetry is further characterized by its 't Hooft anomaly, that is determined by the braiding between two such topological lines, and which is given (on spin manifolds) by the level of the CS theory, $N$. Turning on a background 2-cocycle $B_2$ for the 1-form symmetry on the domain wall, its inflow action is given exactly by the SPT phase \eqref{anomaly usp}.

Note that one could turn the reasoning around and use the knowledge of the SPT phase difference between the two sides of the domain wall to determine the theory that has the right anomaly to absorb it. However, such theory is not unique. In the present case, one would just determine that the domain wall theory has to have a topological subsector coupling to the bulk 1-form symmetry given by the $\mathcal{A}^{2,N}$ theory of \cite{Hsin:2018vcg}. Indeed, $\mathrm{SU}(2)_N$ CS has such a subsector, but is an obviously richer theory. For instance, for $N$ even one would actually think that there is no need for a non-trivial domain wall theory since the SPT does not jump across the domain wall. On the other hand $\mathrm{SU}(2)_N$ CS is a non-trivial TQFT also for $N$ even. Supersymmetry is more powerful in determining the theory living on the domain wall, which in the present case is given by \eqref{dw_sp_1}, because it has additional constraints on the number of vacua that such a gapped theory must have, while the anomaly only gives a constraint on the minimal number of such vacua.

\subsection{Domain walls - holographic description}

Holographically, domain walls are described by 5-branes wrapping compact 3-cycles of the internal geometry, with the other three directions being along $\mathbb{R}^{1,3}$ \cite{Klebanov:2000hb,Maldacena:2000yy,Acharya:2001dz} (see also \cite{Apruzzi:2021phx} for a construction within the KS model).

In our setup these are described by D5-branes wrapping the blown-up 3-cycle $Y_3$ of the orientifolded conifold $X_\epsilon$, that sits in twisted homology.\footnote{The orientifold acts on the string world-sheet. This means that in going around a non-contractible loop, the orientation of the string world-sheet is reversed. The D5 charge is odd under such operation, so the wrapping modes of a D5 are classified by twisted homology \cite{Witten:1998xy}.} They are localized at the apex of $X_\epsilon$ because their tension is minimized there  
(in agreement with domain walls being an IR effect), see Appendix \ref{app: tension of domain walls} for a derivation.\footnote{Note that unlike the string worldsheets that are holographically dual to the Wilson loops, and which never reach $\tau=0$, the D5-branes that are dual to the domain walls do sit at $\tau=0$, where there is a mild curvature singularity. The latter was argued in \cite{Aramini:2025twg} to be the result of the truncation to the zero-mode sector of the backreaction of the orientifold planes. A solution that fully takes into account higher derivative corrections is expected to resolve this singularity. We also expect that the low-energy CS action on the D5-branes, being topological, is robust under the small corrections that lead to such resolution, allowing us to discuss its features irrespectively of the presence of the singularity.} Note that such D5-branes are 1/2-BPS with respect to the D3-brane configuration generating the geometry, in agreement with the field theory result, eq.~\eqref{dw_sp_1}. 
The conifold and the deformed conifold are described, respectively, by the following holomorphic equations in $\mathbb{C}^4$
\begin{equation}
\label{condefcon}
z_1^2 + z_2^2 + z_3^2 + z_4^2 = 0 \ , \quad  z_1^2 + z_2^2 + z_3^2 + z_4^2 = \epsilon^2   \ .
\end{equation}
The blown-up $S^3$ at the tip of the deformed conifold corresponds to the real slice
\begin{equation}
\label{S3def}
\sum_i^4 x_i^2 = \epsilon^2 \quad \mbox{where} \quad z_i = x_i + i y_i \ .
\end{equation}
The orientifold involution acts on the complex coordinates as $z_{1,2,3} \rightarrow z_{1,2,3}$ and $z_4 \rightarrow - z_4$, which  is a symmetry for both the conifold and the deformed conifold.\footnote{Strictly speaking, the geometry of our background is not simply the orientifold of the deformed conifold, since the metric found in \cite{Aramini:2025twg} differs from that of KS. However, since in this section we are only interested in the existence of a blown-up 3-cycle, and since the two backgrounds share the same topology, we will use the deformed conifold as a convenient reference.}

In particular, from eq.~\eqref{S3def} one sees that the blown-up $S^3$ at the tip is invariant as a set under the orientifold projection, with the equator $S^2 \subset S^3$ defined by $x_4=0, x_1^2 + x_2^2 + x_3^2 = \epsilon^2$ being a fixed locus. A D5-brane wrapped on such three-cycle and its image under the orientifold projection have the same support, since the involution is an isometry of the $S^3$. As a result, the gauge group on the D5 world-volume is of $\USp$ type \cite{Gimon:1996rq}. This can be seen by noticing that the action on the Chan-Paton factors is equivalent to the one on the D3-branes supporting the  orientifold gauge theory, see Appendix \ref{app: SUGRA solution}.

Let us take for definiteness $k=1$. Besides the pure SYM term which comes from the DBI action, the world-volume action of a D5-brane  contains also a coupling $\sim \int C_2 \wedge  \Tr\left({\cal F} \wedge {\cal F}\right)$ which provides a non-trivial contribution since the RR 2-form potential is not vanishing at the tip. Here, $\Tr(\cdot\cdot)$ denotes the bilinear invariant form on $\mathfrak{usp}(2)$ normalized such that long roots have length $\sqrt{2}$. Recalling that $F_3 = dC_2 - C_0 \, H_3$ and that $H_3$ vanishes at $\tau=0$ \cite{Aramini:2025twg}, we get that
\begin{align} 
\label{cshol}
\begin{split}
    &  \frac{1}{2} \frac{\mu_5}{2}\int C_2 \wedge  \Tr\left({\cal F} \wedge {\cal F}\right) = 
  - \frac{1}{4}\frac{1}{(2\pi)^5 }\int C_2 \wedge  \Tr \left((2\pi  F) \wedge (2\pi  F)\right) = \\ 
    &  = 
    \frac{1}{8\pi} \frac{1}{(2\pi)^2}\int F_3 \wedge \mathrm{CS}_3(A) = 
    \frac{M_0}{8\pi}\int_{\mathbb{R}^{1,2}}  \mathrm{CS}_3(A) \ , 
    \end{split}
\end{align}
where $\mathrm{CS}_3(A)=\Tr(A\wedge d A + \frac{2}{3}A^3)$ is the usual Chern--Simons term for the three-dimensional $\mathfrak{usp}(2)$-gauge field $A$, with curvature $F$, satisfying $\Tr(F\wedge F) = d\, \mathrm{CS}_3(A)$. In the last step we have used eq.~\eqref{F3_tau0} and the extra factor of 1/2 with respect to the usual normalization of the WZ term arises because the solution  is derived in the covering, unorientifolded space, and fluxes in the physical space should hence be divided by 2 \cite{Aramini:2025twg}. As shown there, the gauge/gravity dictionary relating $M_0$ and the gauge group rank of the pure SYM theory on which the confining vacuum sits, $\USp(2N)$ in the present notation, reads
\begin{equation}
\label{eq: M0 w N}
    M_0 = 2N+2 \ . 
\end{equation}
Therefore, it follows from eq.~\eqref{cshol} that the CS level induced on the 3d effective theory, to which the D5-brane theory reduces to at energies lower than the KK scale of the cycle $Y_3$, is $k=N+1$. 

We thus conclude that the effective 3d theory on a D5-brane wrapped on $Y_3$ is ${\cal N}=1$ YMCS theory with gauge group $\USp(2)$ and CS level $N+1$, in perfect agreement with the field theory answer, eq.~\eqref{purek1_Susy}. 

The $N+1$ vacua of pure $\USp(2N)$ SYM  are distinguished by the phase of the gaugino condensate and are related by the action of the $N+1$ broken generators of the non-anomalous $\mathbb{Z}_{2(N+1)}$ R-symmetry, which is spontaneously broken to $\mathbb{Z}_2$ in the IR 
\begin{equation}
    \langle \lambda \lambda \rangle \sim \Lambda^3 e^{2\pi ik/(N+1)} \ , \quad  k=0,\dots, N \ .
\end{equation}
The phase difference between the $j^{th}$ and the $(j+k)^{th}$ vacua is captured by a $2\pi k$ shift of the $\theta$ angle, which is a symmetry of the theory but, because of the chiral anomaly, has the effect of changing the phase of the gaugino condensate as $\alpha \rightarrow \alpha + 2\pi k/(N+1)$. A $k$-wall should capture such jump \cite{Witten:1998uka}. 

 D5-branes are magnetically charged under $C_2$ so they source $F_3$ through the Bianchi identity. As discussed in \cite{Aramini:2025twg}, the gauge/gravity dictionary relates the $\theta$ angle of the confining gauge group at the bottom of the cascade with the $C_2$-flux as $\theta = 1/(4\pi )\int_{S^2} C_2$.\footnote{Along the cascade the gauge/gravity dictionary is more involved and the $\theta$ angles (one for each gauge group factor) are proportional to the combination  $\int_{S^2} C_2 \pm  C_0 \int_{S^2} B_2$. However, at the bottom of the cascade the $B_2$-flux vanishes and the map simplifies, see eq.~\eqref{F3_tau0}.} This is key to determining the jump in the $\theta$ angle accomplished by a D5-brane wrapped on $Y_3$.

Let us take for definiteness a single D5, which corresponds to a 1-wall. Suppose that it is located at $x_3=0$ in the four-dimensional spacetime. The D5 world-volume is along $W_6 =  \mathbb{R}^{1,2} \times Y_3$. The magnetic $F_3$-flux sourced by the D5-brane is along the non-compact three cycle dual to $Y_3$, locally $B = \mathbb{R}^+ \times S^2$. Then, if one compares the RR flux data on the two sides of the wall, the wrapped D5 contributes as\footnote{In the orientifolded  space the $F_3$-flux is quantized as $\frac{1}{8\pi^2 } \int F_3 \in \mathbb{Z}$ \cite{Aramini:2025twg}.}  
\begin{equation}
   \frac{1}{4\pi} \int_{B, x_3 >0} F_3 - \frac{1}{4\pi}  \int_{B, x_3 <0} F_3 =   2\pi \ . 
\end{equation}
Since $\int_{B} F_3 = \int_{S^2_\infty} C_2$,  
this corresponds to a jump in the SYM $\theta$ angle by $2\pi$, as expected for a domain wall interpolating between two adjacent vacua.

 \section{On strict mass gap in the confining vacuum}
\label{sec: mass_gap}

The confining vacuum we have been considering is at the origin of the mesonic branch of the two-node quiver gauge theory, and no continuous 0-form symmetry is spontaneously broken there: the non-anomalous R-symmetry that is spontaneously broken is discrete, while the $\mathrm{SU}(2)_F$ flavor symmetry is preserved   \cite{Aramini:2025twg}.  Therefore, we do not expect the gravity dual to contain any excitations corresponding to massless modes. 

This should be contrasted with the vacuum described by the KS solution \cite{Klebanov:2000hb}, which is not fully gapped. The KS confining vacuum sits on the baryonic branch and, by Goldstone theorem, admits a massless scalar mode - a full massless chiral superfield, in fact. Its existence has indeed been proven,  holographically, in \cite{Gubser:2004qj,Gubser:2004tf,Argurio:2006my}. The massless pseudo-scalar bound state corresponding to the Goldstone boson of the spontaneously broken baryonic symmetry was found to correspond to a certain perturbation of the RR 2-form potential which also mixes with the RR 4-form potential \cite{Gubser:2004qj}. Its scalar companion has also been found, and corresponds to a massless glueball coming from a mixture of the NS-NS 2-form and a metric deformation. The existence of the pseudo-scalar mode is tied to the very possibility of having tensionful D1-branes extended in four-dimensional spacetime and localized at the tip of the deformed conifold. Such D1-branes do not correspond to ’t Hooft lines (which are screened in the KS theory), but rather to solitonic strings that couple to this massless mode—hence dubbed axion—in that they create a monodromy for the supergravity pseudo-scalar mode.\footnote{In \cite{Argurio:2006my} the fermionic partner of the axion (and of its scalar companion, the saxion), was also identified.}  

The D1-branes of interest are stretched along two of the flat spacetime directions. This suggests, as we explain below, that in our setup they are unstable. By the reasoning above, this would imply that the corresponding massless axionic mode is absent (and, by supersymmetry, the whole chiral superfield), in agreement with the fact that our vacuum is fully gapped. The argument is as follows.

The O9 orientifold of type IIB string theory, namely type I string theory, does not admit stable BPS D3-branes. More precisely, D3-branes are non-BPS and unstable because of an open-string tachyon in the 3-3 sector \cite{Gimon:1996rq,Bergman:2000tm}. Upon performing two T-dualities along the D3-brane worldvolume, this system is mapped to a D1-brane lying on an O7-plane, which by the same reasoning is also unstable, since T-duality is an exact equivalence in string theory. This conclusion is unchanged if the background is not ten-dimensional flat Minkowski space but rather $\mathbb{R}^{1,3}\times$ Calabi–Yau, provided the D3-brane (and hence the D1-brane after T-duality) extends along $\mathbb{R}^{1,3}$. Indeed, the instability of the type I D3-brane is a local open-string effect, so for this specific configuration the geometry of the internal space transverse to the D-brane does not play any essential role. This shows that the axionic string of interest is absent in our setup, as anticipated.

To explicitly check that there are no massless modes in our background, one should analyze the spectrum of all fluctuations around our solution. Concerning the modes that in the KS background are associated to the massless goldstone multiplet of the spontaneously broken baryonic symmetry, it is worth to notice, following \cite{Gubser:2004qj}, that they are odd under the $\mathbb{Z}_2$ space-time symmetry involved in the orientifold action. A complete analysis of the spectrum, however, is beyond the scope of the present paper and is left to future investigations.

\section{Discussion}
\label{sec: disc}

In this paper, we have presented several checks of the holographic duality proposed in \cite{Aramini:2025twg} for a cascading supersymmetric gauge theory that flows from a SCFT in the UV to a discrete set of isolated, fully gapped vacua in the IR. In particular, we computed the expectation value of the fundamental Wilson loop, which is an order parameter for confinement, derived the three-dimensional supersymmetric theory describing the dynamics of domain walls interpolating between the different isolated vacua, and provided convincing arguments for the absence of massless excitations around such vacua. All these results are in excellent agreement with field theory expectations.

There are several additional aspects that one may wish to consider. For one thing, throughout this work we have implicitly assumed the simply connected global form of the gauge group. Since the gauge theory admits a $\mathbb{Z}_2$ one-form symmetry, one may gauge this symmetry, thereby passing to a non-simply connected global form. In this case, at the bottom of the cascade one would then expect the gauge group to be $\USp(2N)/\mathbb{Z}_2$, with its two variants, $(\USp(2N)/\mathbb{Z}_2)_0$ and $(\USp(2N)/\mathbb{Z}_2)_1$. In this case, the Wilson loop in the fundamental representation is no longer a genuine line operator. For $(\USp(2N)/\mathbb{Z}_2)_0$, the relevant genuine probe is the minimal 't Hooft loop, whose magnetic charge is labeled by the weight corresponding to the spinor representation of the Langlands dual group, $\Spin(2N+1)$. For $(\USp(2N)/\mathbb{Z}_2)_1$, instead, the relevant order parameter is the corresponding dyonic line operator with the same magnetic label, but dressed by one unit of electric $\mathbb{Z}_2$ charge. Such order parameters obey a perimeter or an area law depending on the variant, the even or odd nature of $N$, and the vacuum that is being considered \cite{Aharony:2013hda,Argurio:2023lwl}. Following \cite{Witten:1998xy,Bergman:2022otk}, one expects such 't Hooft lines to admit a holographic description in terms of fat strings, namely NS5-branes wrapping appropriate (possibly twisted) four-cycles of the internal space, while the dyonic lines should be described by the same objects bound to a fundamental string. One would similarly like to identify the holographic duals of the topological defect operators that generate the corresponding generalized symmetries under which Wilson, 't Hooft, and dyonic lines are charged.

A powerful framework to classify all possible global forms of a QFT and describe both charged operators and symmetry operators is the Symmetry Topological Field Theory \cite{Gaiotto:2014kfa, Ji:2019eqo, Gaiotto:2020iye, Apruzzi:2021nmk, Freed:2022qnc}. Given a gauge/gravity dual pair, the corresponding SymTFT is expected to admit a holographic description in terms of the topological sector of the effective five-dimensional theory obtained by compactifying type IIB string theory on the internal space $X_5$,  as anticipated in \cite{Witten:1998wy}. 

To address this problem, as well as to properly describe generalized symmetries and charged operators in full-fledged type IIB string theory, a better  understanding of the topology of the internal space $X_\epsilon$ is required. In fact, going beyond our orientifold setup, the need for such an understanding  arises more generally whenever one considers compactifications of type IIB string theory on $AdS_5 \times X_5$ backgrounds, or more generally  on warped $AdS$ geometries, where the internal space $X_5$ has singularities, for example because it is realized as an orbifold or orientifold quotient of a space by an action with fixed loci. These constitute a large class of potentially interesting gauge/gravity duals in the context of D-branes at CY singularities. To the best of our knowledge, virtually no concrete such examples have been treated in a satisfactory way with currently known techniques. It would be highly desirable to identify some appropriate hands-on tools with which to study them, in the spirit of \cite{Witten:1998xy}. Work along these lines is in progress.

\subsection*{Acknowledgements}

We thank Francesco Benini for discussions and Eduardo Garc\'ia-Valdecasas for collaboration on related material. F.A., M.B. and P.M. acknowledge support by INFN Iniziativa Specifica ST\&FI. R.A. is a Research Director of the F.R.S.-FNRS (Belgium). The research of R.A. is supported by IISN-Belgium (convention 4.4503.15) and through an ARC advanced project. 

\appendix
\section{Recap of the gauge theory and the dual supergravity solution}
\label{app: SUGRA solution}

In this appendix, we review some basic aspects of the supersymmetric gauge theory and the corresponding dual type IIB supergravity solution found in \cite{Aramini:2025twg}. We refer the reader to the original paper for more details.

\begin{figure}[t]
    \centering
    \includegraphics[width=0.65\linewidth]{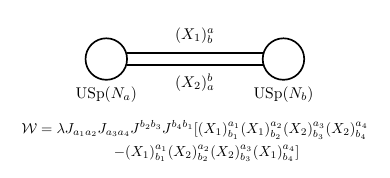}
    \caption{The quiver gauge theory on a stack of regular and fractional branes at the tip of the conifold in presence of an O7-plane. Nodes represent symplectic gauge groups, lines connecting them represent chiral superfields transforming in the fundamental representation of each of the two gauge groups. $X_1,X_2$ transform as a doublet under a $SU(2)$ flavor symmetry. The matrix $J$ is the invariant tensor of the unitary symplectic group.}
    \label{fig:placeholder}
\end{figure}

The setup consists of a stack of regular and fractional D3-branes placed at the tip of the conifold in the presence of an O7-plane. The corresponding gauge theory is an ${\cal N}=1$ quiver gauge theory with gauge group $\USp \times \USp$, two bifundamental chiral multiplets $X_i$, with $i=1,2$, and a quartic superpotential, see figure \ref{fig:placeholder}.  
The theory admits an $SU(2)$ flavor symmetry, a non-ABJ-anomalous discrete R-symmetry, and a $\mathbb{Z}_2^{(1)}$ one-form symmetry.

Much like the KS solution \cite{Klebanov:2000hb}, the RG flow can be described in terms of a Seiberg-duality cascade and interpolates between an SCFT in the UV (more precisely, a one-dimensional conformal manifold) and confinement in the IR. For an appropriate choice of the numbers of regular and fractional branes, the theory reduces, at the origin of the scalar-field VEV moduli space, to ${\cal N}=1$ $\USp(2N)$ SYM, which admits $N+1$ isolated, fully gapped confining vacua. Two salient features make this theory quite different from the KS theory and any of its known relatives: first, the cascade is finite, meaning that the RG flow originates from a SCFT with a finite number of degrees of freedom; second, the confining vacua are fully gapped, in the sense that they sit at the origin of moduli space and do not admit any massless excitations around them. These two properties are the main reason why this construction is particularly interesting.

The dual supergravity solution is obtained by the backreaction of the D3-branes and the O7-plane on the conifold geometry. Since orientifold planes behave similarly to D-branes in their couplings to closed-string fields, the resulting solution exhibits non-trivial profiles for all supergravity fields: the metric, the NSNS $B_2$ field, and all RR potentials. 

The ten-dimensional spacetime coordinates are $\{t,x,y,z, r, \theta_1,\varphi_1,\theta_2,\varphi_2,\psi\}$, where the $x^{\mu}$'s parametrize four-dimensional Minkowski spacetime, and the other six coordinates, with ranges $r \in [0,+\infty)$, $\theta_1,\theta_2 \in [0,\pi]$, $\varphi_1 \sim \varphi_1 + 2\pi$, $\varphi_2 \sim \varphi_2 + 2 \pi$ and $\psi \sim \psi + 4 \pi$,  parametrize the internal space. This set of coordinates is adapted to the conifold geometry, which is the original background before the backreaction of the D3-branes and the O7-plane is taken into account.

The solution for the metric in the Einstein frame reads
\begin{equation}
    \label{eq: metric Ansatz}
    \begin{split}
        ds^2= \, \,&h(r)^{-\frac{1}{2}}dx_{1,3}^2+ h(r)^\frac{1}{2}\bigg[dr^2+e^{2G_1(r)} (\sigma_1^2+\sigma_2^2)+e^{2G_2(r)}((\omega_1+g(r)\sigma_1)^2+\\&(\omega_2+g(r)\sigma_2)^2) 
        +\frac{e^{2G_3(r)}}{9}(\omega_3+\sigma_3)^2
        \bigg] \ ,
    \end{split}
\end{equation}
where $dx_{1,3}^2$ is the Minkowski metric in four-dimensions, and we have defined the following set of one-forms 
\begin{align}
\label{sigmaomega}
    &\sigma_1=d\theta_1 \ , \quad  \sigma_2=\sin{\theta_1}d\varphi_1 \ ,  \quad \sigma_3=\cos{\theta_1}d\varphi_1 \ , \quad \omega_1=\sin{\psi}\sin{\theta_2}d\varphi_2+\cos{\psi}d\theta_2 \nonumber\\
    &\omega_2=-\cos{\psi}\sin{\theta_2}d\varphi_2+\sin{\psi}d\theta_2 \ , \quad \omega_3=d\psi+\cos{\theta_2}d\varphi_2 \ .
\end{align}
This metric has a $\mathbb{Z}_2$ symmetry acting as $\theta_1 \leftrightarrow \theta_2 \ , \ \varphi_1 \leftrightarrow \varphi_2$, which is the spacetime isometry gauged by the orientifold. As in \cite{Aramini:2025twg}, everything is written in the covering space, with the understanding that the physical space is its $\mathbb{Z}_2$ quotient and in units such that $\alpha' =1$.

The functions $h, G_1, G_2, G_3$, and $g$ appearing in the metric are determined by the equations of motion. A clever change of coordinates from $r$ to a new holographic coordinate $\tau$, defined by $d\tau=3e^{-G_3}dr$, gives these functions a rather compact form, which reads
\begin{equation}
    \label{eq: solution deformed conifold}
    \begin{split}
     &  h(\tau)=-\frac{\pi M_0^{\, 2}}{4\mu^\frac{8}{3}}
    \int\limits_{\tau}^{\infty} dx  \, H(x) \  , \quad g^2 = 1 - e^{2(G_1-G_2)} \  , \quad 
    e^{2G_1(\tau)}= \frac{1}{4} \mu^\frac{4}{3} \tilde{\Lambda}(\tau) \frac{\sinh^2\tau}{\cosh\tau} \ , 
    \\
    & e^{2G_2(\tau)} = \frac{1}{4} \mu^\frac{4}{3} \tilde{\Lambda}(\tau) \cosh\tau \  , \quad 
        e^{2G_3(\tau)}= 6 \mu^{\frac{4}{3}} \frac{\tau-\tau_0}{\tilde{\Lambda}^2(\tau)}  \ ,
    \end{split} 
\end{equation}
where $\tau_0, \,\mu$ and $M_0$ are integration constants and 
\begin{equation}
\label{eq: HtildeL}
    \begin{split}
       &H(x) = \frac{x\coth x-1}{(x-\tau_0)^2\sinh^2 x}\times 
    \frac{[-\cosh{(2x)}+4 x(x-\tau_0)+1-(x-2\tau_0)\sinh{(2x)}]}{\left[2\left(\sinh{(2x)}-x\right)(x-\tau_0)-\cosh{(2x)}+2x\tau_0+1\right]^\frac{2}{3}} \ , \\
       & \tilde{\Lambda}(\tau)=\frac{[2(\sinh{(
    2\tau)}-\tau)(\tau-\tau_0)-\cosh{(2\tau)}+2\tau\tau_0+1]^\frac{1}{3}}{\sinh{\tau}} \ .
    \end{split}
\end{equation}
 The metric \eqref{eq: metric Ansatz} resembles that of the deformed conifold.  Sections of the internal space at large values of $\tau$ asymptote to $T^{1,1}\simeq S^2 \times S^3$ with its standard Sasaki--Einstein metric, with volumes of the two spheres that increase with $\tau$. In contrast, as $\tau$ goes to zero, the $S^2$ shrinks to zero size while the $S^3$ reaches a minimal, non-zero volume, exactly as in the KS solution. This is where the internal space terminates. The constants $\mu$ and $\tau_0$ are both related to the volume of the $S^3$ at $\tau = 0$. Finally, $M_0$ is the flux of $F_3$ at $\tau=0$.\footnote{{\color{red}} The solution derived in \cite{Aramini:2025twg} depends on a fourth integration constant, denoted there by $\tau_c$, which corresponds precisely to the scale at which the $S^2$ shrinks to a point and the $S^3$ attains its minimal volume. Here, for convenience, we set $\tau_c=0$.}
 
In contrast to the KS solution, the dilaton has a non-trivial profile. The solution reads
\begin{equation}
\label{eq: dilaton_1}
    e^\phi=\frac{\pi}{2}\frac{1}{\tau-\tau_0} \ .
\end{equation}
At $\tau=\tau_0$, the dilaton diverges. Therefore, for the solution to make sense, the geometry has to end before this value of $\tau$ is reached. With our choice of setting $\tau_c=0$, this implies $\tau_0 <0$. From \eqref{eq: dilaton_1} we see that the maximal value of the string coupling $g_s\sim 1/|\tau_0|$ is reached at $\tau=0$. This implies that the validity of the semi-classical supergravity approximation requires $|\tau_0| \gg 1$. 

In addition to the metric and the dilaton, the behavior of the NSNS and RR fluxes at $\tau =0$ is also relevant to the discussion in Sections \ref{sec: 1-form_WL} and \ref{sec: domain_wall}. This corresponds to the IR regime of the gauge theory. In particular
\begin{equation}
\label{F3_tau0}
     \begin{split}
&     B_2 (\tau = 0) = 0  \ , \\
& M_{eff} (\tau=0) = \frac{1}{4 \pi^2} \int_{S^{3}} F_{3} \, \, \Bigg |_{\tau =0} = M_0 \ , \\
& N_{eff} (\tau=0) = \frac{1}{(4\pi)^2} \int_{T^{1,1}} F_5 \, \, \Bigg |_{\tau =0} = N_0 \ . 
\end{split}
\end{equation}
As explained in \cite{Aramini:2025twg}, the running of $B_2$, together with that of the dilaton, corresponds to the running of the gauge couplings in the dual field theory, while the flux of $F_3$ is related to the difference in the ranks of the two gauge groups along the cascade. The confining vacuum our solution describes sits at the origin of the moduli space of the orientifold gauge theory, where the theory reduces to pure ${\cal N}=1$ SYM with gauge group $\USp(2N)$ in present notation. As shown in \cite{Aramini:2025twg}, this fixes the integration constant $M_0$ to be $M_0 = 2N+2$ and $N_0=0$.

\section{Asymptotic expansions of \texorpdfstring{$S$}{S} at large and small \texorpdfstring{$L$}{L}}
\label{app: asymptotic expansion}

In this appendix we expand the NG action \eqref{eq:S and L} in the limit of large and small $L$, i.e. $\tau_m \to \tau_{\star}$ and $\tau_m\to \infty$, to show that the expectation value of the Wilson loop in the dual gauge theory follows an area law in the IR and a Coulomb law (plus $\log$ corrections) in the UV, as advocated in the main body of the paper. The expressions of $S$ and $L$ obtained in \eqref{eq:S and L} read
\begin{equation}\label{eq:S and L appendix}
\begin{split}
    S &= \frac{T}{3 \pi } \left(\int_{\tau_m}^{\infty} \left( \frac{1}{{\sqrt{1-\frac{h\, e^{\phi_m}}{h_m e^\phi}}}} -1 \right) e^{G_3} e^\frac{\phi}{2}\, d\tau - \int_0^{\tau_m} e^{G_3} e^\frac{\phi}{2}\, d\tau\right)\ , \\
    L &=  \frac{2h_m^\frac{1}{2}}{3 e^\frac{\phi_m}{2}} \int_{\tau_m}^{\infty} \frac{\frac{h\,  e^{\phi_m}}{h_m e^\phi}}{\sqrt{1-\frac{h\,  e^{\phi_m}}{h_m e^\phi}}} e^{G_3} e^\frac{\phi}{2}\, d \tau \ ,
\end{split}
\end{equation}

\subsubsection*{Large $L$ expansion}

As shown in \cref{sec: 1-form_WL}, both $S$ and $L$ diverge as $\tau_m \to \tau_{\star}$. Here, we prove that their ratio $S/L$ converges to a finite constant in the same limit, meaning that fundamental Wilson loops satisfy an area law. Let us first focus on the integral expression for $L$ in \eqref{eq:S and L appendix}. Notice that the integral converges at $\infty$; since we are interested in the singular behavior of $L$ as $\tau_m\to \tau_\star$, we can assume that $\tau_m$ is arbitrarily close to $\tau_\star$ and fix a $UV$ cutoff $\bar\tau$ such that $0\ll \tau_m-\tau_\star \ll \bar\tau-\tau_\star \ll 1$. More explicitly, we consider the integral
\begin{equation}\label{eq: cutoffed L}
    \frac{2h_m^\frac{1}{2}}{3 e^\frac{\phi_m}{2}} \int_{\tau_m}^{\bar\tau} \frac{\frac{h\,  e^{\phi_m}}{h_m e^\phi}}{\sqrt{1-\frac{h\,  e^{\phi_m}}{h_m e^\phi}}} e^{G_3} e^\frac{\phi}{2}\, d \tau \ .
\end{equation}
Expanding the integrand around $\tau_m$, we obtain
\begin{equation}
    \frac{c+\mathcal{O}(\tau-\tau_m)}{\sqrt{a(\tau-\tau_m)+b(\tau-\tau_m)^2+\mathcal{O}((\tau-\tau_m)^3)}}\ ,
\end{equation}
where $a,b$ and $c$ are the coefficients appearing in the Taylor expansions of the denominator and the numerator, respectively, for $\tau$ close to $\tau_m$. 
After the change of coordinates $x=\tau-\tau_m$, denoting $\epsilon=\bar{\tau}-\tau_m$, the divergent part of the integral in \eqref{eq: cutoffed L} takes the form
\begin{equation}
    \int_0^\epsilon \frac{c+\mathcal{O}(x)}{\sqrt{ax+bx^2+\mathcal{O}(x^3)}}\, dx \ .
\end{equation}
A priori, all coefficients in the expansion depend on $\tau_m$. Note that $\tau_{\star}$ is \textit{defined} as the value of $\tau_m$ for which $a(\tau_m)=0$, which causes the integral to diverge. Therefore, 
in order to Taylor-expand the square root in the denominator for $x$ close to zero, in a way which makes sense in the limit $\tau_m \to \tau_{\star}$, it is useful to keep the first two terms as the ``leading order'', since they never vanish together. This expansion yields 
\begin{equation}
    \int_0^\epsilon \left(\frac{c}{\sqrt{ax+bx^2}}+\frac{\mathcal{O}(x)}{\sqrt{ax+bx^2}}\right)\, dx \ .
\end{equation}
Notice that in the limit $\tau_m\to \tau_{\star}$, which implies $a\to0$, the first term diverges, while the second converges. Moreover, we can control this finite contribution by tuning $\epsilon$, which we can take to be arbitrarily small. Hence, the divergence of $L$ as $\tau\to\tau_\star$ can only arise from the constant term in the numerator. This divergent piece can be computed exactly, giving 
\begin{equation}
    \int_0^\epsilon \frac{c}{\sqrt{ax+bx^2}}\, dx =\frac{2c\,\sinh^{-1}\left({\sqrt{\frac{b\epsilon}{a}}}\right)}{\sqrt{b}}\ .
\end{equation}
Note that the definition of $\tau_\star$ implies that the coefficients $a$ and $b$ are positive provided $\tau_m$ is close enough to $\tau_\star$, which is one of our working assumptions in this paragraph. Then, using $a(\tau_m)= q (\tau_m-\tau_{\star})+\smallO(\tau_m-\tau_{\star})$, we find that, in the limit $\tau_m \to \tau_{\star}$,
\begin{equation}
    L(\tau_m) \underset{\tau_m\to\tau_{\star}}{\sim} - \log{(\tau_m-\tau_{\star})} \ .
\end{equation}
One can easily see that it is exactly the same behavior which leads to the divergence of $S(\tau_m)$ as $\tau\to\tau_\star$. Therefore, we conclude that there is a linear relation between $S$ and $L$. Keeping track of numerical factors, one finds that
\begin{equation}
    S \underset{L\to \infty}{\sim} \left(\frac{1}{2\pi}\left(h(\tau_\star)e^{-\phi(\tau_\star)}\right)^{-1/2}\right) T L\ ,
\end{equation}
where the coefficient in parenthesis is then the confining string tension (recall that $h(\tau_\star)e^{-\phi(\tau_\star)}$ is the maximum value of $h(\tau)e^{-\phi(\tau)}$). For large $|\tau_0|$ a simple estimate gives that the string tension is
\begin{equation}
\label{str_tension}
    T_s \sim \frac{\mu^{4/3}|\tau_0|^{1/3}}{M_0} \ .
\end{equation}

\subsubsection*{Small $L$ expansion}

To analyze this regime, we need the large $\tau$ expansion of $h$, $e^{G_3}$ and $e^\phi$. Here and in the following, we will always omit overall constants, which are irrelevant for the present discussion. The expansions are
\begin{equation}\label{eq:expansion h, phi, G3}
    h(\tau) =  \tau^{-\frac{2}{3}} e^{-\frac{4\tau}{3}} (1+\mathcal{O}(1/\tau))\ , \quad e^{G_3} =  \tau^{\frac{1}{6}}e^{\frac{\tau}{3}}(1+\mathcal{O}(1/\tau))\ , \quad e^{\frac{\phi}{2}} = \tau^{-\frac{1}{2}}(1+\mathcal{O}(1/\tau))\ .
\end{equation}
Plugging these expansions into the definition of $L$ in \eqref{eq:S and L appendix}, we can immediately estimate $L$ to be
\begin{equation}
    L = \frac{e^{\frac{\phi_m}{2}}}{h_m^{\frac{1}{2}}}\int_{\tau_m}^{\infty} e^{-\tau}(1+\mathcal{O}(1/\tau))\ d\tau = \frac{e^{\frac{\phi_m}{2}}}{h_m^{\frac{1}{2}}} e^{-\tau_m}(1+\mathcal{O}(1/\tau_m))\ .
\end{equation}
So, for large $\tau_m$, using again \eqref{eq:expansion h, phi, G3}, we find
\begin{equation}\label{eq:L for large tau_m}
    L \underset{\tau_m\to \infty}{\sim} \tau_m^{-\frac{1}{6}}e^{-\frac{\tau_m}{3}}\ .
\end{equation}
Now let us focus on $S$. For $\tau_m \to \infty$, $S$ diverges.
We can consider the two terms in \eqref{eq:S and L appendix} separately. The second term is
\begin{equation}\begin{split}
      & - \int_{0}^{\tau_m} e^{\frac{\phi}{2} + G_3} d \tau \sim - \int_{0}^{\bar \tau} e^{\frac{\phi}{2} + G_3} d \tau - \int_{\bar \tau}^{\tau_m} e^{\frac{\phi}{2} + G_3} d \tau \\
    \sim & - \int_{0}^{\bar \tau} e^{\frac{\phi}{2} + G_3} d \tau - \int_{\bar \tau}^{\tau_m} \tau^{-\frac{1}{3}} e^{\frac{1}{3} \tau} d \tau \sim c(\bar \tau) - 3\tau_m^{-\frac{1}{3}} e^{\frac{1}{3}\tau_m} \sim - 3\tau_m^{-\frac{1}{3}} e^{\frac{1}{3}\tau_m} \, ,
\end{split}\end{equation}
where we have introduced a lower cutoff $\bar{\tau}$ to isolate the divergent part of the integral. \\
One can check that the first term also diverges in the same way, although with the opposite sign. The sum of the two contributions is negative, so that we get
\begin{equation}\label{eq:S for large tau_m}
    S \underset{\tau_m\to  \infty}{\sim} -\tau_m^{-\frac{1}{3}} e^{\frac{1}{3}\tau_m} \, .
\end{equation}
Now we can simply invert \eqref{eq:L for large tau_m} at leading order and plug it into \eqref{eq:S for large tau_m}. This finally gives for $S$
\begin{equation}
    S \underset{L\to 0}{\sim}  - \frac{T}{L\sqrt{\log(1/L)}} \ .
\end{equation}

\section{Tension of the domain walls}
\label{app: tension of domain walls}

The gauge theory domain walls discussed in Section \ref{sec: domain_wall} are realized holographically as D5 branes wrapping the blown-up 3-cycle $Y_3$ of the deformed orientifolded conifold, which in the covering space is an $S^3$ defined $\forall \tau$ by $\theta_2=\text{constant}, \phi_2=\text{constant}$. Here we want to check that the tension of the domain wall is minimized at the tip of the geometry ($\tau=0$). To do so, we need to evaluate the DBI action
\begin{equation}\label{eq:DBI}
    S_{\text{D5}}= \frac{T_5}{2} \int_{\mathbb{R}^{1,2}\times S^3} e^{-\phi}\sqrt{-\det(g+B_2)} \ ,
\end{equation}
where $T_p = \frac{1}{(2 \pi)^p}$ (the extra factor of 1/2 arises because the solution is derived in the covering space and fluxes in the physical space should hence be divided by 2 \cite{Aramini:2025twg}), and the closed string fields $g$ and $B_2$ are pulled back on the D5 world-volume. 

Using eqs.~\eqref{eq: solution deformed conifold} and \eqref{eq: HtildeL}, one finds that the metric on the six-dimensional internal space is
\begin{equation}
    \label{eq: metric deformed conifold}
    ds^2_6=\frac{1}{2}\mu^\frac{4}{3} \tilde{\Lambda}(\tau)\left[\frac{4}{3}\frac{\tau-\tau_0}{\tilde{\Lambda}^3(\tau)}(d\tau^2+(g^5)^2)+ \cosh^2{\left(\frac{\tau}{2}\right)}((g^3)^2+(g^4)^2)+\sinh^2{\left(\frac{\tau}{2}\right)}((g^1)^2+(g^2)^2)\right] ,
\end{equation}
where the $g^i$'s are one-forms defined by
\begin{align} \label{eq: g_i definition}
    g^1&=\frac{1}{\sqrt{2}}(\omega_2-\sigma_2) & g^2&=\frac{1}{\sqrt{2}}(-\omega_1+\sigma_1) & \nonumber
    g^3&=-\frac{1}{\sqrt{2}}(\omega_2+\sigma_2) \\ g^4&=\frac{1}{\sqrt{2}}(\omega_1+\sigma_1) &
    g^5&=\omega_3+\sigma_3 \ .
\end{align}
In this parametrization, the pullback of the metric onto the D5 worldvolume kills the $d \tau^2$, $(g^1)^2$ and $(g^2)^2$ contributions, leaving the metric of a squashed three-sphere. The explicit expression for $B_2$ in the supergravity solution is given by \cite{Aramini:2025twg}
\begin{equation}
    B_2 = \frac{M_0}{2} \Bigl[ f\, g^1 \wedge g^2\,+\,k\, g^3 \wedge 
g^4 \Bigr] \, ,
\end{equation}
where
\begin{equation}
\begin{split}
        f &= \frac{\pi}{2 \, (\tau-\tau_0)}\frac{\tau\coth{\tau} - 1}{2 \sinh {\tau}}(\cosh{\tau}- 1)\ ,\\
        k &= \frac{\pi}{2 \, (\tau-\tau_0)}\frac{\tau\coth{\tau} - 1}{2 \sinh {\tau}}(\cosh{\tau}+ 1)\ \ . 
    \end{split}
\end{equation}
Using these explicit expressions for the metric and the $B_2$ field in \eqref{eq:DBI}, we find
\begin{equation}\begin{split} \label{eq:det}
-\det(g+B_2) &=  \ e^{\frac{3}{2} \phi} h^{-\frac{3}{2}} \biggl(\frac{1}{64} e^{\frac{3}{2} \phi} h^{\frac{3}{2}} \mu^4 p^2(\tau) \, q^4(\tau) \sin^2{\theta_1} + \frac{1}{64} e^{\frac{\phi}{2}} h^{\frac{1}{2}} \mu^{\frac{8}{3}} M_0^2 (f+k)^2 p^2(\tau) \sin^2{\theta_1}  \biggr) \\ & = \frac{1}{64} e^{3 \phi} \mu^{4}  p^2(\tau) \,  q^4 (\tau) \sin^2{\theta_1} \biggl(1 + \frac{M_0^2 (f+k)^2}{e^{\phi} \mu^{\frac{8}{3}} \, h(\tau) \, q^4(\tau)} \biggr) \, ,
\end{split}
\end{equation}
where 
\begin{equation}
    p^2(\tau)=\frac{8}{3}\frac{\tau-\tau_0}{\tilde{\Lambda}^2(\tau)}\ , \quad q^2(\tau)=\tilde{\Lambda}(\tau)\cosh^2{\left(\frac{\tau}{2}\right)}\ .
\end{equation}
Notice that since the D5 spans three directions along Minkowski spacetime and three directions along the $S^3$, the warp factor contribution to the first term in eq.~\eqref{eq:det} cancels exactly. This guarantees that, as $\tau \rightarrow \infty$, the tension of the D5 goes to infinity (due to the hyperbolic cosine in $q(\tau)$), so that the minimum is attained at a finite value of $\tau$. 

Plugging \eqref{eq:det} into \eqref{eq:DBI}, we get
\begin{equation}\begin{split}
    T_{\text{DW}} & = T_5 \, \mu^2 \pi^2 e^{\frac{\phi}{2}}\, p(\tau) \, q^2(\tau) \sqrt{1+\frac{\pi M_0^2}{2 \mu^{\frac{8}{3}} h(\tau)}\frac{(\tau \coth{\tau}-1)^2\coth^2{\tau}}{(\tau-\tau_0) \, q^4(\tau)}} = \\
    & = \frac{2 \sqrt{3}}{3} T_5 \, \mu^2 \pi^{\frac{5}{2}} \cosh^2{\left(\frac{\tau}{2}\right)} \sqrt{1+\frac{\pi M_0^2}{2 \mu^{\frac{8}{3}} h(\tau)}\frac{(\tau \coth{\tau}-1)^2\coth^2{\tau}}{(\tau-\tau_0) \ \tilde{\Lambda}^2(\tau) \, \cosh^4{\left(\frac{\tau}{2}\right)}}} \, .
\end{split}\end{equation}
In this last step, there is a non-trivial cancellation between $e^{\frac{\phi}{2}}$ and $p$: the factors of $\sqrt{\tau-\tau_0}$ exactly cancel, and this is ultimately the reason why the minimum sits at $\tau=0$. More concretely, the correction term inside the square root is $\mathcal{O} (\tau^2)$ for $\tau \rightarrow 0$, which means that
\begin{equation}
\label{eq : Tdw}
   T_{\text{DW}} \underset{\tau\to 0}{\sim}  \mu^2 (1 + c \tau^2) \ ,
\end{equation}
where $c >0$ is a positive constant. This implies that $\tau=0$ is a minimum for $T_{\text{DW}}$. Moreover, since the square root in $T_{\text{DW}}$ is always greater than $1$ and since the hyperbolic cosine is an increasing function, $\tau = 0$ is a global minimum. Therefore, the domain walls on the field theory side will be given by D5 branes wrapping the $Y_3$ at the tip of the orientifolded deformed conifold, as anticipated. Their tension is hence given by $T_{\text{DW}}(0) \sim  \mu^2$. 

On general grounds, one expects that in the holographic limit, once the glueball mass is identified with the strong-coupling scale $\Lambda$, the string tension and the domain wall tension scale as $\Lambda^2$ and $N \Lambda^3$, respectively, possibly up to (model-dependent) powers of $g_s N$, see e.g. \cite{Bertolini:2003iv,Bigazzi:2002gyi}. For example, in the KS model one finds $T_s \sim (g_s N) \Lambda^2$ and $T_{\text{DW}} \sim (g_s N)^2 N \Lambda^3$, meaning that $T_s^3/T_{\text{DW}}^2$ scales as $1/(g_s N^3)$ \cite{Klebanov:2000hb,Herzog:2002ih}. In our case, from eqs.~\eqref{str_tension} and \eqref{eq : Tdw} one gets $T_s^3/T_{\text{DW}}^2 \sim 1/(g_s M_0^3)$, where $g_s \sim 1/|\tau_0|$, see the discussion below eq.~\eqref{eq: dilaton_1}. We then see that the scaling is the same as in the KS model, of which our solution is indeed a close relative.

\bibliographystyle{utphys}
\bibliography{confinement_2}

\end{document}